\documentstyle[prd,preprint,aps]{revtex}
\newcommand{\newc}{\newcommand}
\newc{\gsim}{\lower.7ex\hbox{$\;\stackrel{\textstyle>}{\sim}\;$}}
\newc{\lsim}{\lower.7ex\hbox{$\;\stackrel{\textstyle<}{\sim}\;$}}
\newc{\gev}{\,{\rm GeV}}
\newc{\mev}{\,{\rm MeV}}
\newc{\ev}{\,{\rm eV}}
\newc{\kev}{\,{\rm keV}}
\newc{\tev}{\,{\rm TeV}}
\newc{\mz}{m_Z}
\newc{\mpl}{M_{Pl}}
\newc{\chifc}{\chi_{{}_{\!F\!C}}}
\newc\order{{\cal O}}
\newc\CO{\order}
\newc\CL{{\cal L}}
\newc\CY{{\cal Y}}
\newc\CH{{\cal H}}
\newc\CM{{\cal M}}
\newc\CF{{\cal F}}
\newc\CD{{\cal D}}
\newc\CN{{\cal N}}
\newc{\eps}{\epsilon}
\newc{\re}{\mbox{Re}\,}
\newc{\im}{\mbox{Im}\,}
\newc{\invpb}{\,\mbox{pb}^{-1}}
\newc{\invfb}{\,\mbox{fb}^{-1}}
\newc{\yddiag}{{\bf D}}
\newc{\yddiagd}{{\bf D^\dagger}}
\newc{\yudiag}{{\bf U}}
\newc{\yudiagd}{{\bf U^\dagger}}
\newc{\yd}{{\bf Y_D}}
\newc{\ydd}{{\bf Y_D^\dagger}}
\newc{\yu}{{\bf Y_U}}
\newc{\yud}{{\bf Y_U^\dagger}}
\newc{\ckm}{{\bf V}}
\newc{\ckmd}{{\bf V^\dagger}}
\newc{\ckmz}{{\bf V^0}}
\newc{\ckmzd}{{\bf V^{0\dagger}}}
\newc{\X}{{\bf X}}
\newc{\bbbar}{B^0-\bar B^0}

\newc{\sgn}{\mbox{sgn}\,}
\newc{\m}{{\bf m}}
\newc{\msusy}{M_{\rm SUSY}}
\newc{\munif}{M_{\rm unif}}
\newc{\slepton}{{\tilde\ell}}
\newc{\Slepton}{{\tilde L}}
\newc{\sneutrino}{{\tilde\nu}}
\newc{\selectron}{{\tilde e}}
\newc{\stau}{{\tilde\tau}}
\relax
\def\beq{\begin{equation}}
\def\eeq{\end{equation}}
\def\bea{\begin{eqnarray}}
\def\eea{\end{eqnarray}}
\newc{\ie}{{\it i.e.}}          \newc{\etal}{{\it et al.}}
\newc{\eg}{{\it e.g.}}          \newc{\etc}{{\it etc.}}
\newc{\cf}{{\it c.f.}}
\def\Dsl{\,\raise.15ex\hbox{/}\mkern-13.5mu D} 
\def\delsl{\raise.15ex\hbox{/}\kern-.57em\partial}
\def\Ksl{\hbox{/\kern-.6000em\rm K}}
\def\Asl{\hbox{/\kern-.6500em \rm A}}
\def\Qsl{\hbox{/\kern-.6000em\rm Q}}
\def\gradsl{\hbox{/\kern-.6500em$\nabla$}}
\def\bar#1{\overline{#1}}

\begin{document}
\draft
\title{Measuring the fermionic couplings of the Higgs boson at future
colliders as a probe of a non-minimal flavor structure}
\author{ J.L. D\'{\i}az-Cruz$^{(a,b)}$, R. Noriega-Papaqui$^{(c)}$
and  A. Rosado$^{(a,c)}$}
\address{$^{(a)}$ Cuerpo Acad\'emico de Part\'{\i}culas, Campos y Relatividad
de la BUAP.\\
$^{(b)}$ Facultad de Ciencias F\'{\i}sico-Matem\'aticas, BUAP.\\
Apdo. Postal 1364, C.P. 72000 Puebla, Pue., M\'exico\\
$^{(c)}$ Instituto de F\'{\i}sica, BUAP. Apdo. Postal J-48, C.P. 72570 Puebla, Pue., M\'exico}

\date{\today}
\maketitle
\begin{abstract}
We study the fermionic couplings of the neutral Higgs bosons in the THDM,
assuming a four-texture structure for the Yukawa matrices. We then derive
the low-energy constraints on the model, focusing in b-quark and lepton
physics, and apply them to study Higgs boson detection at future colliders.
We show that the bound on the flavor-violating parameter $\chi_{sb}$ obtained
from the contribution due to the $b \bar{s} h^0$-coupling to the decay
$b \to s + \gamma$ (roughly of the order $10^{-1}-10^{-2}$) is approximately
a factor 10 more restrictive than that obtained from the current bound on
$\Gamma(B^0_s \to \mu^- \mu^+)$ (which gives a bound on $\chi_{sb}$ of the
order $10^0-10^{-1}$), while LFV decay $\mu \to e \gamma$ constraints
$\chi_{\mu\tau} \lsim 10^{-2}$. These constraints imply that a future muon
collider could be able to detect Higgs boson signals from the decays
$h^0 \to \mu^+ \tau^-$ and $h^0 \to b \bar{s}$ for $\tan\beta \lsim 15$,
while such signals turn out to be too small for $\tan\beta \gsim 20$. At a
hadron collider it is further possible to study the Higgs boson coupling
$h^0 b \bar{b}$, by searching for the associated production of the Higgs
boson with $b \bar{b}$ pairs.
\end{abstract}

\setcounter{footnote}{0}
\setcounter{page}{2}
\setcounter{section}{0}
\setcounter{subsection}{0}
\setcounter{subsubsection}{0}

\narrowtext

\section{Introduction.}

Despite the success of the Standard Model (SM) in the gauge and fermion
sectors, the Higgs sector remains the least tested aspect of the model,
and the mechanism of electroweak symmetry breaking (EWSB) is still a puzzle.
However, the analysis of raditive corrections within the SM \cite{hixradc},
points towards the existence of a Higgs boson with mass of the order of the
EW scale, which could be detected in the early stages of LHC \cite{hixphen}.
On the other hand, the SM is often considered
as an effective theory, valid up to an energy scale of $O(TeV)$,
and eventually it will be replaced by a more fundamental theory,
which will explain, among other things, the physics behind EWSB 
and perhaps even the origin of flavor. Several examples of candidate 
theories, which range 
from supersymmetry \cite{susyrev} to deconstruction \cite{deconst},
include a Higgs sector with two scalar doublets, which has 
a rich structure and predicts interesting phenomenology \cite{mssmhix}.  
 The general two-higgs doublet model (THDM) has 
a potential problem with flavor changing neutral currents (FCNC)
mediated by the Higgs bosons, which arises when each quark type (u and d) 
is allowed to couple to both Higgs doublets, and FCNC could be induced at
large rates that may jeopardize the model. 
The possible solutions to this problem of the THDM 
involve an assumption about the Yukawa structure of the model. 
To discuss them it is convenient to refer to the Yukawa lagrangian, which is
written for the quarks fields as follows:
\begin{equation}
{\cal{L}}_Y = Y^{u}_1\bar{Q}_L \Phi_{1} u_{R} + 
                   Y^{u}_2 \bar{Q}_L\Phi_{2}u_{R} +
Y^{d}_1\bar{Q}_L \Phi_{1} d_{R} + Y^{d}_2 \bar{Q}_L\Phi_{2}d_{R} 
\end{equation}
where $\Phi_{1,2}=(\phi^+_{1,2}, \phi^0_{1,2})^T$ denote the Higgs doublets.
The specific choices for the Yukawa matrices $Y^q_{1,2}$ ($q=u,d$) define
the versions of the THDM known as I, II or III, which involve
the following mechanisms, that are aimed either to eliminate the otherwise
unbearable FCNC problem or at least to keep it under control, namely:

\begin{enumerate}
\item {\it{DISCRETE SYMMETRIES.}}
A discrete symmetry can be invoked to allow a given fermion
type (u or d-quarks for instance) to couple to a single Higgs
doublet, and in such case FCNC are absent at tree-level. 
In particular, when a single Higgs field gives masses to both types 
of quarks (either $Y^u_1=Y^d_1=0$ or $Y^u_2=Y^d_2=0$), the resulting
model is referred as THDM-I. On the other hand, when each type of quark 
couples to a different Higgs doublet (either $Y^u_1=Y^d_2=0$ or 
$Y^u_2=Y^d_1=0$), the model is known as the THDM-II.
 This THDM-II pattern is highly motivated because it arises at tree-level 
in the minimal SUSY extension for the SM (MSSM) \cite{mssmhix}.

\item {\it{RADIATIVE SUPRESSION.}} 
When each fermion type couples to both Higgs doublets, 
FCNC could be kept under control if there exists
a hierarchy between $Y^{u,d}_1$ and $Y^{u,d}_2$.
Namely, a given set of Yukawa matrices is present at tree-level,
but the other ones  arise only as a radiative effect. 
This occurs for instance in the MSSM, where the 
type-II THDM structure is not protected by any symmetry, 
and is transformed into a type-III THDM (see bellow), through 
the loop effects of sfermions and gauginos. 
Namely, the Yukawa couplings that are already present at tree-level
in the MSSM ($Y^d_1, Y^u_2$) receive radiative corrections, while the terms
($Y^d_2, Y^u_1$) are induced at one-loop level. 

 In particular, when the ``seesaw'' mechanism~\cite{seesaw} is 
implemented in the MSSM to explain the observed neutrino 
masses ~\cite{atmospheric,solar}, lepton flavor violation (LFV) 
appears naturally in the right-handed neutrino sector, which is then 
communicated to the sleptons and from there to the charged leptons 
and Higgs sector. These corrections allow the neutral 
Higgs bosons to mediate LFV, in particular  it was found 
that the (Higgs-mediated) tau decay $\tau \to 3\mu$ \cite{bakotau}
as well as the (real) Higgs boson decay $H\to \tau\mu$ \cite{myhlfvA}, 
can enter into possible detection domain. 
Similar effects are known to arise in the quark sector,
for instance $B\to\mu\mu$ can reach branching fractions 
at large $\tan\beta$, that can be probed at Run~II of the 
Tevatron~\cite{prlbako,bmuanalysis}. 

\item {\it{FLAVOR SYMMETRIES.}}
Suppression for FCNC can also be achived when a certain form of
the Yukawa matrices that reproduce the observed fermion masses and
mixing angles is implemented in the model, which is then named as 
THDM-III.  This could be done either by implementing the Frogart-Nielsen 
mechanism to generate the fermion mass hierarchies \cite{FN},
or by studying a certain ansatz for the fermion mass matrices 
\cite{fritzsch}. The first proposal for the Higgs boson couplings
along these lines was posed
in \cite{chengsher,others}, it was based on the six-texture 
form of the mass matrices, namely:
\begin{displaymath}
M_l= 
\left( \begin{array}{ccc}
0 & C_{q} & 0 \\
C_{q}^{*} & 0 & B_{q} \\
0 & B_{q}^{*} & A_{q}
\end{array}\right).
\end{displaymath}
Then, by assuming that each Yukawa matrix $Y^q_{1,2}$ has the same 
hierarchy, one finds: $A_{q}\simeq m_{q_3}$, $B_{q}\simeq
\sqrt{m_{q_2}m_{q_3}}$ and
$C_{q}\simeq \sqrt{m_{q_1}m_{q_2}}$. Then, the fermion-fermion$'$-Higgs boson
($f f' \phi^0$) couplings obey the following pattern:
$Hf_{i}f_{j} \sim \sqrt{m_{f_i}m_{f_j}} / m_{W}$, 
which is known as the Cheng-Sher ansatz. This brings under control
the FCNC problem, and it has been extensively studied in the literature to
search for flavor-violating signals in the Higgs sector \cite{muchos}.

\end{enumerate}

In this paper we are interested in studying the THDM-III.
However, the six-texture ansatz seems disfavored by current data
on the CKM mixing angles. More recently, mass matrices with
four-texture ansatz have been considered, and are found in better 
agreement with the observed data \cite{fourtext,xing}. It is interesting
then to investigate how the Cheng-Sher form of the $f f' \phi^0$
couplings, get modified when one replaces the six-texture matrices 
by the four-texture ansatz. This paper is aimed precisely to study
this question; we want to derive the form of the $f  f' \phi^0$
couplings and to discuss how and when the resulting predictions could be 
tested, both in rare quark and lepton decays and in the phenomenology of 
the Higgs bosons \cite{myhlfvA}. Unlike previous studies, we
keep in our analysis the effect of the complex phases, which modify the
FCNC Higgs boson couplings.

The organization of the paper goes as follows: In section 2, we discuss the
lagrangian for the THDM with the four-texture form for the mass matrices, and
present the results for the $f  f' \phi^0$ vertices in the quark sector.
Then, in section 3 we study the constraints impossed on the
parameters of the model from low energy flavor violating processes. Section 4
includes the predictions of the model for both, flavor conserving (FC) and
flavor violating (FV) Higgs boson decays. While in section 5, we discuss the
capabilities of future $\mu^+ \mu^-$ and hadron colliders to detect such
decays. Finally, section 6 contains our conclusions.

\section{The quark sector of the THDM-III with four-texture mass matrices}

The Yukawa lagrangian of the THDM-III is written for the quarks fields as
follows:
\begin{equation}
{\cal{L}}_Y^q = Y^{u}_1\bar{Q}_L \Phi_{1} u_{R} + 
                   Y^{u}_2 \bar{Q}_L\Phi_{2}u_{R} +
Y^{d}_1\bar{Q}_L \Phi_{1} d_{R} + Y^{d}_2 \bar{Q}_L\Phi_{2}d_{R} 
\end{equation}
where $\Phi_{1,2}=(\phi^+_{1,2}, \phi^0_{1,2})^T$ denote the Higgs doublets.
The specific choices for the Yukawa matrices $Y^q_{1,2}$ ($q=u,d$) define
the versions of the THDM known as I, II or III.

After spontaneous symmetry breaking the quark mass matrix is given by,
\begin{equation}
M_q= \frac{1}{\sqrt{2}}(v_{1}Y_{1}^{q}+v_{2}Y_{2}^{q}),
\end{equation}

We will asuume that both Yukawa matrices $Y^q_1$ and $Y^q_2$ have the
four-texture form and Hermitic; following the conventions of \cite{fourtext}, the
quark mass matrix is then written as:

\begin{displaymath}
M_q= 
\left( \begin{array}{ccc}
0 & C_{q} & 0 \\
C_{q}^{*} & \tilde{B}_{q} & B_{q} \\
0 & B_{q}^{*} & A_{q}
\end{array}\right).
\end{displaymath}
when $\tilde{B}_{q}\to 0$ one recovers the six-texture form.
We also consider the hierarchy: \\
$\mid A_{q}\mid \, \gg \, \mid \tilde{B}_{q}\mid,\mid B_{q}\mid ,\mid C_{q}\mid$,
which is supported by the observed fermion masses in the SM.

Because of the hermicity condition, both $\tilde{B}_{q}$ and
$A_{q}$ are real parameters, while the phases of $C_q$ and $B_q$,
$\Phi_{B_q,C_q}$, can be removed from the mass matrix $M_q$
by defining: $M_q=P_q^\dagger \tilde{M}_q P_q$, where 
$P_q=diag[1, e^{i\Phi_{C_q}},  e^{i(\Phi_{B_q}+\Phi_{C_q})}]$, 
and the mass matrix $\tilde{M}_q$ includes only the real parts
of $M_q$. The  diagonalization of $\tilde{M}_q$ is then obtained by an
orthogonal matrix
$O_q$, such that the diagonal mass matrix is: 
$\bar{M}_{q} = O_q^{T}\tilde{M}_{q}O_q$.

\bigskip

The lagrangian (2) can be expanded in terms of the mass-eigenstates for
the neutral ($h^0,H^0,A^0$) and charged Higgs bosons ($H^\pm$). The
interactions of the neutral Higgs bosons with the d-type and u-type are
given by ($u,u'=u,c,t.$ and $d,d\,'=d,s,b.$),

\begin{eqnarray}
{\cal{L}}_Y^{q} & = & \frac{g}{2}\left(\frac{m_d}{m_W}\right)
\bar{d}\left[\frac{ \, \cos\alpha}{\cos\beta}\delta_{dd'}+ 
\frac{\sqrt{2} \, \sin(\alpha - \beta)}{g \, \cos\beta}
\left(\frac{m_W}{m_d}\right)(\tilde{Y}_2^d)_{dd'}\right]d\,'H^{0} 
\nonumber \\
                 &  &+ \frac{g}{2}\left(\frac{m_d}{m_W}\right)\bar{d}
\left[-\frac{\sin\alpha}{\cos\beta} \delta_{dd'}+  
\frac{\sqrt{2} \, \cos(\alpha - \beta)}{g \, \cos\beta}
\left(\frac{m_W}{m_d}\right)(\tilde{Y}_2^d)_{dd'}\right]d\,' h^{0}
\nonumber \\
                 & &+ \frac{ig}{2}\left(\frac{m_d}{m_W}\right)\bar{d}
\left[-\tan\beta \delta_{dd'}+  \frac{\sqrt{2} }{g \, \cos\beta}
\left(\frac{m_W}{m_d}\right)(\tilde{Y}_2^d)_{dd'}\right]
\gamma^{5}}d\,' A^{0 \nonumber \\
                 & &+ \frac{g}{2}\left(\frac{m_u}{m_W}\right)
\bar{u}\left[\frac{ \, \sin\alpha}{\sin\beta}\delta_{uu'}- 
\frac{\sqrt{2} \, \sin(\alpha - \beta)}{g \, \sin\beta}
\left(\frac{m_W}{m_u}\right)(\tilde{Y}_2^u)_{uu'}\right]u'H^{0} 
\nonumber \\
                 &  &+ \frac{g}{2}\left(\frac{m_u}{m_W}\right)\bar{u}
\left[\frac{\cos\alpha}{\sin\beta} \delta_{uu'}-  
\frac{\sqrt{2} \, \cos(\alpha - \beta)}{g \, \sin\beta}
\left(\frac{m_W}{m_u}\right)(\tilde{Y}_2^u)_{uu'}\right]u' h^{0}
\nonumber \\
                 & &+ \frac{ig}{2}\left(\frac{m_u}{m_W}\right)\bar{u}
\left[-\cot\beta \delta_{uu'} + \frac{\sqrt{2} }{g \, \sin\beta}
\left(\frac{m_W}{m_u}\right)(\tilde{Y}_2^u)_{uu'}\right]
\gamma^{5}}u' A^{0.
\end{eqnarray}
The first term, proportional to $\delta_{qq'}$ corresponds to the
modification of the THDM-II over the SM result, while the term
proportional to $\tilde{Y}_2^q$ denotes the new contribution from
THDM-III. Thus, the $f f' \phi^0$ couplings respect CP-invariance,
despite the fact that the Yukawa matrices include complex phases;
this follows because of the Hermiticity conditions impossed on both
$Y_1^q$ and $Y_2^q$.

The corrections to the quark flavor conserving (FC) and flavor violating
(FV) couplings, depend on the rotated matrix:  $\tilde{Y}_{2}^{q} =
O_q^{T}P_qY_{2}^{q}P_q^\dagger O_q$. We will evaluate $\tilde{Y}_{2}^{q}$
assuming that $Y_2^q$ has a four-texture form, namely:

\begin{equation}
Y_{2}^{q}  =
\left( \begin{array}{ccc}
0 & C_2^q & 0 \\
C_2^{q*} & \tilde{B}_2^q & B_2^q \\
0 & B_2^{q*} & A_2^q
\end{array}\right), \qquad
\mid A_2^q\mid \, \gg \, \mid \tilde{B}_2^q\mid,\mid B_2^q\mid ,\mid
C_2^q\mid.
\end{equation}

The matrix that diagonalizes the real matrix
$\tilde{M}_{q}$ with the four-texture form, is given by:

\begin{displaymath}
O_q =
\left( \begin{array}{ccc}
\sqrt{\frac{\lambda^q_{2}\lambda^q_{3}(A_q-\lambda^q_{1})}{A_q(\lambda^q_{2}-\lambda^q_{1})
(\lambda^q_{3}-\lambda^q_{1})}}& \eta_q \sqrt{\frac{\lambda^q_{1}\lambda^q_{3}
(\lambda^q_{2}-A_q)}{A_q(\lambda^q_{2}-\lambda^q_{1})(\lambda^q_{3}-\lambda^q_{2})}}
& \sqrt{\frac{\lambda^q_{1}\lambda^q_{2}(A_q-\lambda^q_{3})}{A_q(\lambda^q_{3}-
\lambda^q_{1})(\lambda^q_{3}-\lambda^q_{2})}} \\
-\eta_q \sqrt{\frac{\lambda^q_{1}(\lambda^q_{1}-A_q)}{(\lambda^q_{2}-\lambda^q_{1})
(\lambda^q_{3}-\lambda^q_{1})}} & \sqrt{\frac{\lambda^q_{2}(A_q-\lambda^q_{2})}
{(\lambda^q_{2}-\lambda^q_{1})(\lambda^q_{3}-\lambda^q_{2})}} & \sqrt{
\frac{\lambda^q_{3}(\lambda^q_{3}-A_q)}{(\lambda^q_{3}-\lambda^q_{1})(\lambda^q_{3}-
\lambda^q_{2})}} \\
\eta_q \sqrt{\frac{\lambda^q_{1}(A_q-\lambda^q_{2})(A_q-\lambda^q_{3})}{A_q(\lambda^q_{2}
-\lambda^q_{1})(\lambda^q_{3}-\lambda^q_{1})}} & -\sqrt{\frac{\lambda^q_{2}(A_q
-\lambda^q_{1})(\lambda^q_{3}-A_q)}{A_q(\lambda^q_{2}-\lambda^q_{1})(\lambda^q_{3}
-\lambda^q_{2})}} & \sqrt{\frac{\lambda^q_{3}(A_q-\lambda^q_{1})(A_q-\lambda^q_{2})}
{A_q(\lambda^q_{3}-\lambda^q_{1})(\lambda^q_{3}-\lambda^q_{2})}}
\end{array}\right),
\end{displaymath}
where $m^q_1 = \mid \lambda^q _1\mid$, $m^q_2 = \mid \lambda^q _2\mid$,
$m^q_3 = \mid \lambda^q _3\mid$, and $\eta_q = \lambda^q_2/ m^q_2$ $(q=u,d)$.
With $m_u= m^u_1$, $m_c= m^u_2$, and $m_t= m^u_3$; $m_d= m^d_1$, $m_s= m^d_2$,
 and $m_b= m^d_3$.
  
Then the rotated form $\tilde {Y}_2^q$ has the general form,

\begin{eqnarray}
\tilde {Y}_2^q  & = & O_q^TP_qY_{2}^qP_q^{\dagger}O_q \nonumber \\
& = &\left( \begin{array}{ccc}
(\tilde {Y}_2^q)_{11}   & (\tilde {Y}_2^q)_{12}   & (\tilde {Y}_2^q)_{13}   \\
(\tilde {Y}_2^q)_{21}   & (\tilde {Y}_2^q)_{22}   & (\tilde {Y}_2^q)_{23}  \\
(\tilde {Y}_2^q)_{31}   & (\tilde {Y}_2^q)_{32}   & (\tilde {Y}_2^q)_{33}
\end{array}\right).
\end{eqnarray}

However, the full expressions for the resulting elements have a complicated 
form, as it can be appreciated, for instance, by looking at the element  
$(\tilde{Y}_{2}^q)_{22}$, which is displayed here:

\begin{eqnarray}
(\tilde{Y}_2^q)_{22} &=& \eta_q [C^{q*}_2 e^{i\Phi_{C_q}} +C^q_2
e^{-i\Phi_{C_q}}]
\frac{(A_q-\lambda^q_{2})}{m^q_3-\lambda^q_2 } \sqrt{\frac{m^q_1 m^q_3 }{A_q m^q_2}} +
 \tilde{B}^q_2 \frac{A_q-\lambda^q_2}{ m^q_3-\lambda^q_2 }\nonumber \\
& & + A^q_2 \frac{A_q-\lambda^q_2}{ m^q_3-\lambda^q_2 }
- [B^{q*}_2 e^{i\Phi_{B_q}} + B^q_2 e^{-i\Phi_{B_q}}]
\sqrt{\frac{(A_q-\lambda^q_{2})(m^q_3-A_q) } {m^q_3- \lambda^q_2}} 
\end{eqnarray}
where we have taken the limits: $|A_q|, m^q_3, m^q_2 \gg m^q_1$.
The free-parameters are: $\tilde{B^q_{2}}, B^q_{2}, A^q_{2}, A_q$. 

To derive a better suited approximation, we will consider the elements of
the Yukawa matrix $Y_2^l$ as having the same hierarchy as the full mass
matrix, namely:
 
\begin{eqnarray}
C^q_{2} & = &  c^q_{2}\sqrt{\frac{m^q_{1}m^q_{2}m^q_{3}}{A_q}}  \\
B^q_{2} & = &  b^q_{2}\sqrt{(A_q - \lambda^q_{2})(m^q_{3}-A_q)}  \\
\tilde{B}^q_{2} & = & \tilde{b}^q_{2}(m^q_{3}-A_q + \lambda^q_{2})  \\
A^q_{2} & = & a^q_{2}A_q.
\end{eqnarray}

Then, in order to keep the same hierarchy for the elements of the mass 
matrix, we find that $A_q$ must fall within the interval $ (m^q_3- m^q_2)
\leq A_q \leq m^q_3$. Thus, we propose the following relation for $A_q$:

\begin{equation}
A_q  = m^q_{3}(1 -\beta_q z_q),
\end{equation} 
where $z_q = m^q_{2}/m^q_{3} \ll 1$  and $0 \leq \beta_q \leq 1$.

Then, we introduce the matrix $\tilde{\chi}^q$ as follows:

\begin{eqnarray}
\left( \tilde {Y}_2^q \right)_{ij}
&=& \frac{\sqrt{m^q_i m^q_j}}{v} \, \tilde{\chi}^q_{ij} \nonumber\\
&=&\frac{\sqrt{m^q_i m^q_j}}{v}\, {\chi}^q_{ij} \, e^{i \vartheta^q_{ij}}
\end{eqnarray}
which differs from the usual Cheng-Sher ansatz not only because of the
appearence of the complex phases, but also in the form of the real parts
${\chi}^q_{ij} = |\tilde{\chi}^q_{ij}|$.

Expanding in powers of $z_q$, one finds that the elements of the matrix
$\tilde{\chi}^q$ have the following general expressions:

\begin{eqnarray}
\tilde{\chi}^q_{11} & = &  
[\tilde{b}^q_2-(c^{q*}_2e^{i\Phi_{C_q}} +c^q_2e^{-i\Phi_{C_q}} )]\eta_q 
    +[a^q_2+\tilde{b}^q_2-(b^{q*}_2e^{i\Phi_{B_q}} + b^q_2e^{-i\Phi_{B_q}} )]
         \beta_q \nonumber \\
\tilde{\chi}^q_{12} & = & (c^q_2e^{-i\Phi_{C_q}}-\tilde{b}^q_2) -\eta_q[a^q_2+
\tilde{b}^q_2-(b^{q*}_2e^{i\Phi_{B_q}} + b^q_2e^{-i\Phi_{B_q}} )] \beta_q
\nonumber \\
\tilde{\chi}^q_{13} & = & (a^q_2-b^q_2e^{-i\Phi_{B_q}}) \eta_q \sqrt{\beta_q}
                           \nonumber  \\
\tilde{\chi}^q_{22}  & = & \tilde{b}^q_2\eta_q
+[a^q_2+\tilde{b}^q_2-(b^{q*}_2e^{i\Phi_{B_q}} +b^q_2e^{-i\Phi_{B_q}} )]
         \beta_q \nonumber \\
\tilde{\chi}^q_{23} & = & (b^q_2e^{-i\Phi_{B_q}}-a^q_2)
                              \sqrt{\beta_q} \nonumber  \\
\tilde{\chi}^q_{33} & = & a^q_2
\end{eqnarray}

While the diagonal elements $\tilde{\chi}^q_{ii}$ are real, we notice
(Eqs. 14) the appearance of the phases in the off-diagonal elements,
which are essentially unconstrained by present low-energy phenomena.
As we will see next, these phases modify the pattern of flavor violation
in the Higgs sector. For instance, while the Cheng-Sher ansatz predicts
that the LFV couplings $(\tilde{Y}_2^q)_{13}$ and $(\tilde{Y}_2^q)_{23}$
vanish when $a_2^q = b_2^q$, in our case this is no longer valid for
$\cos\Phi_{B_q} \neq 1$. Furthermore the LFV couplings satisfy several
relations, such as: $|\tilde{\chi}^q_{23}| = |\tilde{\chi}^q_{13}|$,
which simplifies the parameter analysis.

In order to perform our phenomenological study we find convenient
to rewrite the lagrangian given in Eq. (4) in terms of the
$\tilde{\chi}_{qq'} = \tilde{\chi}^q_{ij}$ as follows:
\begin{eqnarray}
{\cal{L}}_Y^{q} & = & \frac{g}{2} \, \bar{d}
\left[\left( \, \frac{m_d}{m_W}\right)\frac{\cos\alpha}{\cos\beta} \,
\delta_{dd'} + \frac{\sin(\alpha - \beta)}{\sqrt{2} \, \cos\beta}
\left(\frac{\sqrt{m_d m_{d'}}}{m_W}\right)\tilde{\chi}_{dd'}\right]d\,'H^{0} 
\nonumber \\
                &   & + \frac{g}{2} \, \bar{d}
\left[-\left(\frac{m_d}{m_W}\right)\frac{\sin\alpha}{\cos\beta} \,
\delta_{dd'} + \frac{\cos(\alpha - \beta)}{\sqrt{2} \, \cos\beta}
\left(\frac{\sqrt{m_d m_{d'}}}{m_W}\right)\tilde{\chi}_{dd'}\right]d\,'h^{0}
\nonumber \\
                &   & + \frac{ig}{2} \, \bar{d}
\left[-\left(\frac{m_d}{m_W}\right)\tan\beta \, \delta_{dd'} +
\frac{1}{\sqrt{2} \, \cos\beta}
\left(\frac{\sqrt{m_d m_{d'}}}{m_W}\right)\tilde{\chi}_{dd'}\right]
\gamma^{5}} d\,' A^{0.\nonumber \\
                &   & \frac{g}{2} \, \bar{u}
\left[\left( \, \frac{m_u}{m_W}\right)\frac{\sin\alpha}{\sin\beta} \,
\delta_{uu'} - \frac{\sin(\alpha - \beta)}{\sqrt{2} \, \sin\beta}
\left(\frac{\sqrt{m_u m_{u'}}}{m_W}\right)\tilde{\chi}_{uu'}\right]u'H^{0} 
\nonumber \\
                &   & + \frac{g}{2} \, \bar{u}
\left[\left(\frac{m_u}{m_W}\right)\frac{\cos\alpha}{\sin\beta} \,
\delta_{uu'} - \frac{\cos(\alpha - \beta)}{\sqrt{2} \, \sin\beta}
\left(\frac{\sqrt{m_u m_{u'}}}{m_W}\right)\tilde{\chi}_{uu'}\right]u'h^{0}
\nonumber \\
                &   & + \frac{ig}{2} \, \bar{u}
\left[-\left(\frac{m_u}{m_W}\right)\cot\beta \, \delta_{uu'} +
\frac{1}{\sqrt{2} \, \sin\beta}
\left(\frac{\sqrt{m_u m_{u'}}}{m_W}\right)\tilde{\chi}_{uu'}\right]
\gamma^{5}} u' A^{0.
\end{eqnarray}
where $u,u'=u,c,t.$ and $d,d\,'=d,s,b.$, and unlike the Cheng-Sher
ansatz, $\tilde{\chi}_{qq'}$ $(q \neq q\,')$ are complex.

Finally, for completeness we display here the corresponding lagrangian for
the charged lepton sector, which has been already reported in our previous
work \cite{papaqui}, namely.

\begin{eqnarray}
{\cal{L}}_Y^{l} & = & \frac{g}{2} \, \bar{l}
\left[\left( \, \frac{m_{l}}{m_W}\right)\frac{\cos\alpha}{\cos\beta} \,
\delta_{ll'} + \frac{\sin(\alpha - \beta)}{\sqrt{2} \, \cos\beta}
\left(\frac{\sqrt{m_l m_{l'}}}{m_W}\right)\tilde{\chi}_{ll'}\right]l' H^{0} 
\nonumber \\
                &   & + \frac{g}{2} \, \bar{l}
\left[-\left(\frac{m_{l}}{m_W}\right)\frac{\sin\alpha}{\cos\beta} \,
\delta_{ll'} + \frac{\cos(\alpha - \beta)}{\sqrt{2} \, \cos\beta}
\left(\frac{\sqrt{m_l m_{l'}}}{m_W}\right)\tilde{\chi}_{ll'}\right]l' h^{0}
\nonumber \\
                &   & + \frac{ig}{2} \, \bar{l}
\left[-\left(\frac{m_{l}}{m_W}\right)\tan\beta \, \delta_{ll'} +
\frac{1}{\sqrt{2} \, \cos\beta}
\left(\frac{\sqrt{m_l m_{l'}}}{m_W}\right)\tilde{\chi}_{ll'}\right]
\gamma^{5}}l' A^{0.
\end{eqnarray}
where $l, l'=e, \mu, \tau$.

On the other hand, one can also relate our results with the SUSY-induced
THDM-III, for instance by considering the effective Lagrangian for the
couplings of the charged leptons to the neutral Higgs fields, namely:
\beq
-{\cal L}=\bar{L}_L Y_l l_{R} \phi_1^0 +\bar{L}_L Y_l \left(\eps_1{\bf 1}
+\eps_2 Y_\nu^\dagger Y_\nu\right) l_{R} \phi_2^{0*} + h.c.
\label{FCLag}
\eeq
In this language, LFV results from our inability to simultaneously
diagonalize the term $Y_l$ and the non-holomorphic loop corrections, 
$\eps_2 Y_l Y_\nu^\dagger Y_\nu$. Thus, since the charged lepton masses
cannot be diagonalized in the same basis as their Higgs boson couplings, this
will allow neutral Higgs bosons to mediate LFV processes with rates
proportional to $\eps_2^2$. In terms of our previous notation we have:
$\tilde{Y}_2 = \eps_2 Y_l {Y_\nu}^\dagger {Y_\nu}$. Thus, our result will
cover (for some specific choices of parameters) the general expectations for
the corrections arising in the MSSM.

\section{Bounds on the Flavor Violating Higgs parameters}

Constrains on the FV-Higgs interaction can be obtained by studying
FV transitions. In this section we consider the radiative decay
$b \to s \, \gamma$ and the decay $B^0_s \to \mu^- \mu^+$, which together
with LFV bounds derived in \cite{papaqui} constraint the parameter space
of THDM-III, and determine possible Higgs boson signals that may be detected
at future colliders.

\bigskip

\noindent {\bf 3.1 Radiative decay $b \to s \, \gamma$}. We will make an
estimation of the contribution due to the flavor-violating $f f' \phi^0$
couplings to the standard model branching ratio of $b \to s \, \gamma$ as
follows

\begin{equation}
\Delta Br(b \to s \, \gamma) = \Delta\Gamma(b \to s \, \gamma) \times
\left ( \displaystyle{\sum_{l=e,\mu,\tau}\Gamma(b \to c \, l \,
\bar{\nu}_l)} \right ) ^{-1}
\end{equation}
Such contribution to the branching ratio of
$b \to s \, \gamma$ at one loop level is then given by \cite{chang1}
\begin{eqnarray}
\Delta Br(b \to s \, \gamma) &=& \frac{\alpha_{em} \, m_s \, m^3_b
\, \cos^2(\alpha-\beta)}{16 \, \pi \, m^4_{h^0} \, |V_{cb}|^2
\, \cos^4 \beta } \,\chi_{sb}^2\nonumber\\
& & \times \left |-\sin\alpha + \frac{\cos(\alpha-\beta)}{\sqrt2} \, \tilde{\chi}_{bb}
\right|^2 \, \left| \ln \frac{m^2_b}{m^2_{h^0}}+\frac{3}{2} \right|^2
\end{eqnarray}
From Eqs. (14) we have $\chi_{sb}
= \chi_{db} = |(a^d_2 - b^d_2 e^{-i\Phi_{B_d}})|\sqrt{\beta_d}$. We will
make use of the good agreement between the current experimental value for
$Br(b \to s \, \gamma) = (3.3 \pm 0.4) \times 10^{-4}$ and the theoretical
value obtained for $Br(b \to s \, \gamma) = (3.29 \pm 0.33) \times 10^{-4}$
in the context of the standard model \cite{partdata} to constraint any new
contribution to $Br(b \to s \, \gamma)$, namely
$\Delta Br(b \to s \, \gamma) \lsim 10^{-5}$,
and hence to bound $\chi_{sb} (= \chi_{db})$ as a function of $m_{h^0}$,
$\tilde{\chi}_{bb}$, $\alpha$ and $\tan\beta$.

\noindent a) Assuming $m_{h^0} = 120$ $GeV$ and
$\chi_{bb} = 0$, we depict in Fig. 1 the values of the upper bound
on $\chi_{sb}$ ($(\chi_{sb})_{u.\,b.}^{b \to s \, \gamma}$) as a
function of $\tan\beta$, for $\alpha = \beta, \, \beta - \pi /4, \,
\beta - \pi /3$. 

\noindent b) Taking $\alpha = \beta - \pi /4$ and $\chi_{bb} = 0$, we plot
in Fig. 2 the results for $(\chi_{sb})_{u.\,b.}^{b \to s \, \gamma}$ as a
function of $\tan\beta$, for $m_{h^0} = 80$ $GeV$, 120 $GeV$, 160 $GeV$.

\noindent c) We show in Fig. 3, taking $m_{h^0} = 120$ $GeV$ and
$\alpha = \beta - \pi /4$, our numerical results for
$(\chi_{sb})_{u.\,b.}^{b \to s \, \gamma}$ as a function of real values
\footnote{We will study the dependency on the phases $\vartheta^f_{ij}$
($\tilde{\chi}^f_{ij} = \chi^f_{ij}e^{\vartheta^f_{ij}}$) of the Higgs
phenomenology in a forthcoming paper.} of $\tilde{\chi}_{bb}$ for
$\tan\beta = 5$, 25, 50.

\bigskip

From Figs. 1-3, we conclude that the upper bound on the LFV parameter
$\chi_{sb}$, from the radiative decay $b \to s + \gamma$ measurements,
is much more restrictive for large values of $\tan\beta$,
$\tilde{\chi}_{bb} \sim -1$, $m_{h^0} \approx 80$ $GeV$ and
$\alpha \approx \beta$  . However, one can still say that at the present
time the coupling $\chi_{sb}$ is not highly constrained when
$\tan\beta \sim 5 - 10$, or even for larger values of $\tan\beta$ provided
that $\tilde{\chi}_{bb} \to +1$ or $\alpha \to \beta - \pi/2$, thus
$\tilde{\chi}_{sb}$ could induce interesting direct LFV Higgs boson signals
at future colliders.

\bigskip

\noindent {\bf 3.2 $B^0_{s} \to \mu^- \mu^+$ decay}. The formula to
calculate the width of the decay $B^0_{s} \to \mu^- \mu^+$ at the one
loop level is given as follows \cite{prlbako}
\begin{eqnarray}
\Gamma(B^0_{s} \to \mu^- \mu^+) &=& \frac{G_F^2 \, \eta_{_{QCD}} \, m_B^3 \, f_B^2
\, m_s \, m_b \, m_{\mu}^2 \, \cos^2(\alpha-\beta)}{128 \, \pi \, m^4_{h^0}
\, \cos^4 \beta } \,\chi_{sb}^2\nonumber\\
& & \times \left |-\sin\alpha + \frac{\cos(\alpha-\beta)}{\sqrt2} \,
\tilde{\chi}_{\mu\mu} \right|^2 
\end{eqnarray}
where $G_F= 1.16639^{-5} \, GeV^{-2}$, $\eta_{_{QCD}}\approx 1.5$,
$m_B \simeq 5 \, GeV$, and $f_B=180 \, MeV$.

We will make use of the current experimental limit for
$\Gamma(B^0_{s} \to \mu^- \mu^+) < 8.7 \times 10^{-19}$ $GeV$
\cite{prlbako,abe1} to constraint the LFV parameter $\chi_{sb} (= \chi_{db})$
and the resulting upper bound will be shown as function of $m_{h^0}$,
$\tilde{\chi}_{\mu\mu}$, $\alpha$ and $\tan\beta$.

\noindent a) Assuming $m_{h^0} = 120$ $GeV$ and
$\chi_{\mu\mu} = 0$, we depict in Fig. 4 the values of the upper bound
on $\chi_{sb}$ ($(\chi_{sb})_{u.\,b.}^{B^0_s \to \mu \mu}$) as a
function of $\tan\beta$, for $\alpha = \beta, \, \beta - \pi /4, \,
\beta - \pi /3$. 

\noindent b) Taking $\alpha = \beta - \pi /4$ and $\chi_{\mu\mu} = 0$, we plot
in Fig. 5 the results for $(\chi_{sb})_{u.\,b.}^{B^0_s \to \mu \mu}$ as a
function of $\tan\beta$, for $m_{h^0} = 80$ $GeV$, 120 $GeV$, 160 $GeV$.

\noindent c) We show in Fig. 6, taking $m_{h^0} = 120$ $GeV$ and
$\alpha = \beta - \pi /4$, our numerical results for
$(\chi_{sb})_{u.\,b.}^{B^0_s \to \mu \mu}$ as a function of real values $^1$
of $\tilde{\chi}_{\mu\mu}$ for $\tan\beta = 5$, 25, 50.

\bigskip

From Figs. 4-6, we conclude that the upper bound on the LFV parameter
$\chi_{sb}$, obtained from the experimental bound for the width of the
radiative decay $B^0_s \to \mu^- \mu^+$,
is more restrictive for large values of $\tan\beta$,
$\tilde{\chi}_{\mu\mu} \sim -1$, $m_{h^0} \approx 80$ $GeV$ and
$\alpha \approx \beta$. However, one can still say again that at the present
time the coupling $\chi_{sb}$ is not highly constrained for
$\tan\beta \sim 5 - 10$, or even larger values of $\tan\beta$ provided
that $\tilde{\chi}_{\mu\mu} \to +1$ or $\alpha \to \beta - \pi/2$.

\bigskip

From Eqs. (18) and (19), we obtain the following relation

\begin{eqnarray}
\Gamma(B^0_{s} \to \mu^- \mu^+) &=& \frac{1.22 \times 10^{-14} \, GeV}
{\left| \ln \frac{m^2_b}{m^2_{h^0}}+\frac{3}{2} \right|^2} \,
\frac{\left |-\sin\alpha + \frac{\cos(\alpha-\beta)}{\sqrt2}\tilde{\chi}_
{\mu\mu} \right|^2}
{\left |-\sin\alpha + \frac{\cos(\alpha-\beta)}{\sqrt2}\tilde{\chi}_{bb}
\right|^2} \, \Delta Br(b \to s \, \gamma)
\end{eqnarray}

Assuming that $\tilde{\chi}_{\mu\mu} = \tilde{\chi}_{bb}$ (or
$\chi_{\mu\mu} \lsim 10^{-2}$ and $\chi_{bb} \lsim 10^{-2}$)
and taking $\Delta Br(b \to s \, \gamma) < 10^{-5}$, which is a
conservative bound \cite{partdata}, we get

\begin{eqnarray}
\Gamma(B^0_{s} \to \mu^- \mu^+)&<&\left \{ \begin{array}{llll}
& 6.7 \times 10^{-21} \, GeV &\hspace{0.5in} \mbox{for} \;\; m_{h^0}=80
\, GeV\\
& 4.8 \times 10^{-21} \, GeV &\hspace{0.5in} \mbox{for} \;\; m_{h^0}=120
\, GeV\\
& 3.8 \times 10^{-21} \, GeV &\hspace{0.5in} \mbox{for} \;\; m_{h^0}=160
\, GeV\\
\end{array}
\right.
\end{eqnarray}

\bigskip

Thus, we conclude from (22) that the bound on the parameter $\chi_{sb}$
obtained from the constraint on the contribution due to the
$b \bar{s} h^0$-coupling to the theoretical branching ratio of the radiative
decay $b \to s + \gamma$ is approximately a factor ten more restrictive than
that one obtained from the current experimental bound for
$\Gamma(B^0_s \to \mu^- \mu^+)$ already mentioned \cite{prlbako,abe1}.

\section{Higgs boson decays in the THDM-III}

One of the distinctive characteristic of the SM Higgs boson is the fact that
its coupling to other particle is proportional to the mass of that particle,
which in turn determines the search strategies proposed so far to detect it
at future colliders. In particular, the decay pattern of the Higgs boson is
dominated by the heaviest particle allowed to appear in its decay products.
When one considers extensions of the SM it is important to study possible
deviations from the SM decay pattern as it could provide a method to
discriminate among the different models \cite{lista1}.

Within the context of the THDM-III, which we have been studying, not only
modification of the Higgs boson couplings are predicted, but also the appearance
of new channels with flavor violation, both in the quark and leptonic
sectors \cite{myhlfvA,lista3}.

To explore the characteristics of Higgs boson decays in the THDM-III, we
will focus on the lightest CP-even state ($h^0$), which could be detected
first at LHC. The light Higgs boson-fermion couplings are given by Eqs. (15)
and (16), where we have separated the SM from the corrections that appear in
a THDM-III. In fact, we have also separated the factors that arise in the
THDM-III too. We notice that the correction to the SM result, depends on
$\tan\beta$, $\alpha$ (the mixing angle in the neutral CP-even Higgs sector)
and the factors $\tilde{\chi}_{ij}$ that induce FCNC transitions (for
$i \neq j$) and further corrections to the SM vertex.

In what follows, we will include the decay widths for all the modes that are
allowed kinematically for a Higgs boson with a mass in the range $80 \, GeV
< m_{h^0} < 160 \, GeV$. Namely, we study the branching ratios for the decays
$h^0 \to b\bar{b},\,c\bar{c},\,\tau\bar{\tau},\,\mu\bar{\mu}$ and the
flavor-violating $h^0 \to b\bar{s}(s\bar{b}),\,\tau\bar{\mu}(\mu\bar{\tau})$,
as well as the decays into pairs of gauge bosons with one real an the other
one virtual, {\it i.e.} $h^0 \to WW^*, ZZ^*$.

Making use of Eqs. (15) and (16) we obtain
\begin{eqnarray}
\Gamma(h^0 \to d \bar{d})&=&3\,
\frac{g^2 \, m_{h^0} \, m^2_d}{32 \, \pi \, m^2_W}
\left | -\frac{\sin\alpha }{\cos\beta}+\frac{\cos(\alpha-\beta)}{\sqrt{2}
\cos\beta} \, \tilde{\chi}_{dd} \right |^2 \,
\left(\frac{\lambda(m^2_d,m^2_d,m^2_{h^0})}{m^4_{h^0}} \right)^{3/2}
\end{eqnarray}

\begin{eqnarray}
\Gamma(h^0 \to d \bar{d'})&=&3 \, 
\frac{g^2 \, m_{h^0} \, m_{d} \, m_{d'}}{64 \, \pi \, m^2_W}
\frac{\cos^2(\alpha-\beta)}{\cos^2\beta}
\left(\frac{\lambda(m^2_{d},m^2_{d'},m^2_{h^0})}{m^4_{h^0}}
\right)^{3/2} \chi_{dd'}^2
\end{eqnarray}

\begin{eqnarray}
\Gamma(h^0 \to u \bar{u})&=&3\,
\frac{g^2 \, m_{h^0} \, m^2_u}{32 \, \pi \, m^2_W}
\left | \frac{\cos\alpha }{\sin\beta}-\frac{\cos(\alpha-\beta)}{\sqrt{2}
\sin\beta} \, \tilde{\chi}_{uu} \right |^2 \,
\left(\frac{\lambda(m^2_u,m^2_u,m^2_{h^0})}{m^4_{h^0}} \right)^{3/2}
\end{eqnarray}

\begin{eqnarray}
\Gamma(h^0 \to u \bar{u'})&=&3 \, 
\frac{g^2 \, m_{h^0} \, m_{u} \, m_{u'}}{64 \, \pi \, m^2_W}
\frac{\cos^2(\alpha-\beta)}{\sin^2\beta}
\left(\frac{\lambda(m^2_{u},m^2_{u'},m^2_{h^0})}{m^4_{h^0}}
\right)^{3/2} \chi_{uu'}^2
\end{eqnarray}

\begin{eqnarray}
\Gamma(h^0 \to l \bar{l})&=&
\frac{g^2 \, m_{h^0} \, m^2_{l}}{32 \, \pi \, m^2_W}
\left | -\frac{\sin\alpha }{\cos\beta}+\frac{\cos(\alpha-\beta)}{\sqrt{2}
\cos\beta} \, \tilde{\chi}_{ll} \right |^2 \,
\left(\frac{\lambda(m^2_{l},m^2_{l},m^2_{h^0})}{m^4_{h^0}} \right)^{3/2}
\end{eqnarray}

\begin{eqnarray}
\Gamma(h^0 \to l \bar{l'})&=&
\frac{g^2 \, m_{h^0} \, m_{l} \, m_{l'}}{64 \, \pi \, m^2_W}
\frac{\cos^2(\alpha-\beta)}{\cos^2\beta}
\left(\frac{\lambda(m^2_{l},m^2_{l'},m^2_{h^0})}
{m^4_{h^0}} \right)^{3/2} \chi_{ll'}^2
\end{eqnarray}
where: $\lambda(x,y,z) = (x-y-z)^2 - 4yz$; $u,\,u'=u,\,c,\,t$; $d,\,d\,'=
d\,,s\,,b$; and $l,\,l'=e^-,\, \mu^-,\, \tau^-.$

\noindent For the decays $h^0 \to WW^*, ZZ^*$ we use the corresponding expressions
given in Ref.\cite{hixphen}.

We calculate the branching ratios for all the relevant decay modes that
are allowed kinematically in the range $80 \, GeV < m_{h^0} < 160 \, GeV$;
taking $\alpha = \beta - 3 \pi/8$; assuming $\tilde{\chi}_{ij} = 0.1$ for
$i = j$ and $i \neq j$. We consider the following cases $\tan\beta = 2$,
2.61, 5, 15 and 50.
Our results are displayed in Figs. 7-11, where we notice the important effect
that the factor $\tilde{\chi}_{bb}$ has on the mode $h^0 \to b\bar{b}$,
which could be dominant for certain range of parameters, but it could be
suppressed for other choices. Fig. 12 clarifies what is going on, it shows
the region in the plane $(\alpha-\beta)$ - $\tan\beta$, where the coupling
$h^0 b \bar{b}$ vanishes, and one can notice that this happens even for small
values of the parameter $\tilde{\chi}_{bb}$ ($\approx 0.01$).

We also notice in Figs. 7-11 that the Br for the FCNC mode
$h^0 \to b\bar{s}(\bar{b}s)$
reaches values above $10^{-4}$ and the LFV mode $h^0 \to \tau\bar{\mu}
(\bar{\tau}\mu)$ reaches values above $10^{-5}$ for $5 \lsim \tan\beta
\lsim 50$ and $80 \, GeV \lsim m_{h^0} \lsim 155 \, GeV$. Further, in the
mass range when Br($h^0 \to b\bar{b}$) is not dominant, we find that
the modes $h^0 \to WW^*,\, ZZ^*$ become the dominant ones. 

Overall, our results show that the usual search strategies to look for the
SM Higgs boson in this mass range, may need to be modified in order to cover
the full parameter space of the THDM-III.

In the coming sections we will discuss how the Higgs boson signals could be
searched at a future $\mu^+ \mu^-$-collider. We will also study the reach in parameter space that
could be obtained through the Higgs boson production in association with
a pair of $b$-quarks at LHC, which was found to be relevant in the large
$\tan\beta$ limit for the MSSM \cite{loren3}.

\section{Probing the fermionic Higgs boson couplings at future colliders}

In order to probe the Higgs vertices we will consider first the search
for the LFV Higgs boson decays at future muon colliders, which was proposed some
time ago \cite{mumucolla}, namely we will evaluate the reaction $\mu^-\mu^+
\to h^0 \to f' f''$. Then we will consider the production of Higgs bosons
at the LHC, to probe both LFV and $h^0 b \bar{b}$ couplings.

\bigskip

\noindent {\bf 5.1 Tests of LFV/FCNC Higgs boson couplings at $\mu^- \mu^+$
-colliders.} An option to search for LFV $f f' \phi^0$ couplings, could be
provided by the reaction:
$\mu^-(p_a) + \mu^+(p_b) \to \phi^0 \to f'(p_c) + \bar{f''}(p_d)$.
The $s$-channel Higgs boson cross section (on resonance) is given by:
\begin{equation}
\sigma_{\phi^0}(\mu^-\mu^+ \to f'\bar{f''}) = 4 \pi \frac{\Gamma(\phi^0 \to
\mu^+ \mu^-) \, \Gamma(\phi^0 \to f'\bar{f''})}{(s-m^2_{\phi^0})^2 +
m^2_{\phi^0}\,(\Gamma^{\phi^0}_{tot})^2}
\end{equation}
where $\phi^0$ denotes a neutral Higgs boson which decays to a final state
$f'\bar{f''}$.
The effective cross section $\bar{\sigma}_{\phi^0}$ is obtained by convoluting
with the Gaussian distribution in $\sqrt{s}$ \cite{bbgh1}:
\begin{equation}
\bar{\sigma}_{\phi^0}(\mu^-\mu^+ \to f'\bar{f''}) \simeq \frac{4 \pi}
{m^2_{\phi^0}} \frac{Br(\phi^0 \to \mu^+ \mu^-) \, Br(\phi^0 \to f'\bar{f''})}
{[1+\frac{8}{\pi}(\frac{\sigma_{\sqrt{s}}}{\Gamma^{\phi^0}_{tot}})^2]^{1/2}}
\end{equation}      
$\sigma_{\sqrt{s}}$ can be expressed in terms of the root-mean-square
(rms) Gaussian spread of the energy of an individual beam, $R$, as follows:
\begin{equation}
\sigma_{\sqrt{s}} = (2 \, MeV) \left( \frac{R}{0.003\%} \right)
\left( \frac{\sqrt{s}}{100 \, GeV} \right).
\end{equation}

In this work, we will restrict our numerical
analysis to the case of the light neutral scalar {\it i.e.} $\phi^0=h^0$,
and for the most relevant cases $f'f''= \tau^- \mu^+ (\tau^+ \mu^-), \,
b \bar{s} \, (\bar{b} s)$.

The calculation of $\bar{\sigma}_{h^0}$ requires the evaluation of the
following quantities: $\Gamma(h^0 \to \tau^- \mu^+)$,
$\Gamma(h^0 \to b \bar{s})$, $\Gamma(h^0 \to \mu^- \mu^+)$,
and $\Gamma^{h^0}_{tot}$, which are given in Eqs. (22)-(27).

By performing a detailed numerical analysis one can show that \footnote{We will discuss the
dependency on the parameters $\chi^f_{ij}$ and the phases $\vartheta^f_{ij}$
($\tilde{\chi}^f_{ij} = \chi^f_{ij}e^{\vartheta^f_{ij}}$) of the decay
widths $\Gamma(\phi^0 \to f_i \bar{f}_j)$ for $\phi^0 = h^0$, $H^0$, and
$A^0$ in a forthcoming paper.} 
\begin{equation}
0.98 < \frac{\Gamma(h^0 \to \mu^- \mu^+)}
{\Gamma(h^0 \to \mu^- \mu^+)|_{\tilde{\chi}_{\mu\mu}=0}} < 1.02
\end{equation}
provided that: $80 \, GeV \leq m_{h^0} \leq 160 \, GeV$;
$-\pi/3 \leq \alpha-\beta \leq 0$; $5 \leq \tan\beta \leq 50$;
$|\tilde{\chi}_{\mu\mu}| \lsim 0.01$. Hence, under the previous conditions
we have
\begin{eqnarray}
\Gamma(h^0 \to \mu^- \mu^+) &\simeq&
\Gamma(h^0 \to \mu^- \mu^+) |_{\tilde{\chi}_{\mu\mu}=0} \nonumber\\
&=&\frac{g^2 \, m_{h^0} \, m^2_{\mu}}{32 \, \pi \, m^2_W}
\frac{\sin^2\alpha }{\cos^2\beta} \,
\left( 1-4 \, \frac{m^2_{\mu}}{m^2_{h^0}} \right)^{3/2}
\end{eqnarray}

Assuming $80 \, GeV \leq m_{h^0} \leq 160 \, GeV$, we can write 
\begin{equation}
\Gamma^{h^0}_{tot} =
\sum_{f'} \Gamma(h^0 \to f'\bar{f'}) +
\sum_{f',f''} \Gamma(h^0 \to f'\bar{f''}) +
\Gamma(h^0 \to W W^*) + \Gamma(h^0 \to Z Z^*),
\end{equation}
where $f',f'' \neq t$-quark.

It is also possible to show numerically that $^2$
\begin{equation}
0.98 < \frac{\Gamma^{h^0}_{tot}}
{\Gamma^{h^0}_{tot} |_{\tilde{\chi}_{f'f''}=0}} < 1.06,
\end{equation}
provided that the following conditions are satisfied:
$-\pi/3 \leq \alpha-\beta \leq 0$;
$5 \leq \tan\beta \leq 50$; $|\tilde{\chi}_{ff}| \lsim 0.01$;
$|\tilde{\chi}_{f'f''}| \lsim 1 \; \; (f' \neq f'')$. Hence, under the
previous conditions we can approximate
\begin{equation}
\Gamma^{h^0}_{tot} \simeq
\Gamma^{h^0}_{tot} |_{\tilde{\chi}_{f'f''}=0}.
\end{equation}

We can write the cross-sections of the processes
$\mu^- \mu^+ \to \tau^- \mu^+$ and $\mu^- \mu^+ \to b \, \bar{s}$ as follows:
\begin{equation}
\bar{\sigma}_{h^0} (\mu^- \mu^+ \to \tau^- \mu^+)
\simeq \frac{4 \pi}{m^2_{h^0}}
\frac{Br(h^0 \to \mu^+ \mu^-) \,
Br(h^0 \to \tau^- \mu^+)}
{[1+\frac{8}{\pi}(\frac{\sigma_{\sqrt{s}}}{\Gamma^{h^0}_{tot}})^2]^{1/2}}
\end{equation}      

\begin{equation}
\bar{\sigma}_{h^0} (\mu^- \mu^+ \to b \, \bar{s}) \simeq
\frac{4 \pi}{m^2_{h^0}} \frac{Br(h^0 \to \mu^+ \mu^-) \,Br(h^0 \to b \,
\bar{s})}
{[1+\frac{8}{\pi}(\frac{\sigma_{\sqrt{s}}}{\Gamma^{h^0}_{tot}})^2]^{1/2}}
\end{equation}      
where 
\begin{eqnarray}
Br(h^0 \to \mu^- \mu^+) &\simeq&
\frac{\Gamma(h^0 \to \mu^- \mu^+) |_{\tilde{\chi}_{\mu\mu}=0}}
{\Gamma^{h^0}_{tot} |_{\tilde{\chi}_{f'f''}=0}}\nonumber\\
Br(h^0 \to \tau^- \mu^+) &\simeq&
\frac{\Gamma(h^0 \to \tau^- \mu^+)}
{\Gamma^{h^0}_{tot} |_{\tilde{\chi}_{f'f''}=0}}\nonumber\\
Br(h^0 \to b \, \bar{s}) &\simeq&
\frac{\Gamma(h^0 \to b \, \bar{s})}
{\Gamma^{h^0}_{tot}|_{\tilde{\chi}_{f'f''}=0}}
\end{eqnarray}
provided that $|\tilde{\chi}_{f'f''}| \lsim 10^{-2}$ for
$f'=f''$; and
$|\tilde{\chi}_{f'f''}| \lsim 1$ for $f' \neq f''$.

We will calculate the number of events $\tau^-\mu^+$ $(\tau^+\mu^-)$ produced
in a $\mu^-\mu^+$-collider
\begin{equation}
N^{\mu\mu\to\tau\mu}=
\bar{\sigma}_{h^0}(\mu^- \mu^+ \to \tau^- \mu^+) \times L_{year}.
\end{equation}
\noindent Then our numerical results for $N^{\mu\mu\to\tau\mu}
(s=m^2_{h^0}, \chi_{\mu\tau})$ are shown in Figs. 13-22, as a function of
$\tan\beta$,
by taking: (i) $\chi_{\mu\tau} = 1$ (Figs. 13-17) and (ii) $\chi_{\mu\tau}=
(\chi_{\mu\tau})_{u.\,b.}^{\mu \to e \gamma}$,
the value of the upper bound on $\chi_{\mu\tau}$ obtained from the
experimental measurement of the radiative decay $\mu^+ \to e^+ \gamma$
\cite{papaqui}, we will take the current experimental result
$Br(\mu^+ \to e^+ \gamma) < 1.2 \times 10^{-11}$ \cite{partdata}
(Figs. 18-22). We plot curves for $m_{h^0}=80$ $GeV$, 120 $GeV$, 160 $GeV$,
taking $\alpha=\beta$, $\beta-\pi/4$, $\beta-\pi/3$,
assuming yearly integrated luminosities $L_{year} = 0.1, \, 0.22, \, 1 \,
fb^{-1}$ for beam energy resolutions of $R = 0.003\%, \, 0.01\%, \, 0.1\%$,
respectively \cite{mumucolla}.

From Figs. 13-17, we could expect the production of $\sim 10^1 - 10^2$
$\tau^- \mu^+(\tau^+ \mu^-)$ pairs with a $\mu^-\mu^+$-collider. However,
if we calculate the number of such events using the constraint on
$\chi_{\mu\tau}$ obtained from the experimental bound on the branching
ratio of the LFV process $\mu^+ \to e^+ \, \gamma$, the production rates
are drastically reduced, specially for large values of $\tan\beta$
($\gsim 15$), as it can be observed in Figs. 18-22. We can
conclude that the detection of $\tau^- \mu^+$ or $\tau^+ \mu^-$ events would
be possible for $\tan\beta \lsim 15$, but not for
$\tan\beta \gsim 15$.

On the other hand, the nonobservation of at least an event $\tau^-\mu^+$
(or $\tau^+\mu^-$) in a year would imply that
\begin{equation}
N^{\mu\mu\to\tau\mu}(s=m^2_{h^0},\chi_{\mu\tau}) < 1,
\end{equation}
which would also allow us to put an upper bound on $\chi_{\mu\tau}$, namely:
\begin{equation}
(\chi_{\mu\tau})_{u.\,b.}^{\mu\mu \to \tau\mu}(s=m^2_{h^0}) =
\left [N^{\mu\mu\to\tau\mu}(s=m^2_{h^0},\chi_{\mu\tau} = 1) \right ]^{-1/2}
\end{equation}
According to Figs. 18-22, the $\mu^- \mu^+$  collider measurements could
improve the bound on $\chi_{\mu\tau}$ obtained from the radiative decay
$\mu^+ \to e^+ \, \gamma$, $(\chi_{\mu\tau})_{u.\,b.}^{\mu \to e \gamma}$,
only if $\tan\beta \lsim 15$.

Then, for the quark signals, we will calculate the number of events
$b\,\bar{s}$ $(\bar{b}\,s)$ produced in a $\mu^-\mu^+$-collider, given by:

\begin{equation}
N^{\mu\mu\to b s}=
\bar{\sigma}_{h^0}(\mu^- \mu^+ \to b \bar{s}) \times L_{year}.
\end{equation}
\noindent We depict our numerical results for $N^{\mu\mu\to b s}(s=m^2_{h^0},
\chi_{sb})$ in Figs. 23-32, as a function of $\tan\beta$, by taking:
i) $\chi_{sb} = 1$ (Figs. 23-27) and ii) $\chi_{sb}=(\chi_{sb})_
{u.\,b.}^{b \to s \gamma}$,
the value for the upper bound on $\chi_{sb}$ obtained in subsection 3.1
from the good agreement between the experimental and theoretical value
of the radiative decay $b \to s \gamma$ (Figs. 28-32). We plot curves
for $m_{h^0}=80$ $GeV$, 120 $GeV$, 160 $GeV$, taking
$\alpha=\beta$, $\beta-\pi/4$, $\beta-\pi/3$,
assuming yearly integrated luminosities $L_{year} = 0.1, \, 0.22, \, 1 \,
fb^{-1}$ for beam energy resolutions of $R = 0.003\%, \, 0.01\%, \, 0.1\%$,
respectively \cite{mumucolla}.

From Figs. 23-27, we would expect the production of $\sim 10^2-10^3$
$b \, \bar{s} (\bar{b} \, s)$ pairs at a $\mu^- \mu^+$-collider. However,
the number of such events obtained by using the constraint on $\chi_{sb}$
impossed by the branching ratio of the process $b \to s \, \gamma$, are
drastically reduced for $\tan\beta \gsim 15$, as it can be observed in Figs.
28-32. Again, we can conclude that the detection of $b \, \bar{s}$ or
$b \, \bar{s}$ events would be possible for $\tan\beta \lsim 15$, but not
for $\tan\beta \gsim 15$.

Similarly, the nonobservation of at least an event of the type $b \bar{s}$
(or $\bar{b} s$) in a year, could be used to improve the bound on $\chi_{sb}$
obtained from the radiative decay $b \to s \, \gamma$,
$(\chi_{sb})_{u.\,b.}^{b \to s \gamma}$, only if $\tan\beta \lsim 15$.

\noindent {\bf 5.2 Search for Higgs boson in associated production with
$b$-quarks pairs at LHC.}
The associated production of the Higgs boson in association with a quark pair
$b \bar{b}$, has been found useful to detect the neutral Higgs bosons of the
MSSM \cite{mssmhix}, especially in the large-$\tan\beta$ domain. Here we will
show that this reaction can be also useful to constrain the coupling
$h^0 b \bar{b}$ in the THDM-III.

As shown in Ref.\cite{loren3}, the reaction $pp \to h^0 (\to b \bar{b}) +
b \bar{b} + X$ produces a large sample of events which could be detectable
provided a $K$-factor is above a certain value, which depends on the Higgs
boson mass and the coupling $h^0 b \bar{b}$ (which enter in the event rate
both from the Higgs boson production and decay), this factor is defined as
\begin{equation}
K = \frac{(g_{h^0 b \bar{b}})_{{}_{THDM-III}}}
{(g_{\phi^0 b \bar{b}})_{{}_{SM}}} \, \sqrt{Br(h^0 \to b \bar{b})}\\
\end{equation}
To have a detectable signal at LHC for $m_{h^0} = 150 \, GeV$, the modulus of
this factor has to be above $|K|_{min} = 1.93$, as obtained from a detailed
analysis of signal and backgrounds performed in Ref.\cite{loren3}, to which
we refer for details of kinematical cuts, acceptances and parton
distributions.

In Figs. 33-35, we show the region of the plane $\tan\beta$ - ($\alpha-\beta$),
where the signal for $m_{h^0} = 150 \, GeV$ is detectable. One can notice
that the effect of the parameter $\tilde{\chi}_{b b}$, even for small
values, can have a dramatic impact on the extension of the region of
parameters where the signal is detectable. Therefore, LHC will be able
to constrain the presence of a non-minimal flavor structure (which is
reflected on the parameters $\tilde{\chi}_{ij}$), and provide a decisive
test of the fermionic coupling of the Higgs boson.

\noindent {\bf 5.3 Search for LFV Higgs boson decays at Hadron colliders.}
We will concentrate here on the LFV Higgs boson decays $\phi_i \to \tau \mu$, 
which has a very small  branching ratio within the context of the SM with
light neutrinos ($\lsim 10^{-7}-10^{-8}$ ), so that this channel becomes 
an excellent window for probing new physics\,\cite{myhlfvA,myhlfvB,otherlfvh}.
The decay width for the procces $\phi_i \to \tau \mu$ 
(adding both final states $\tau^+ \mu^-$ and $\tau^- \mu^+$ )
can be written in terms of the decay width $\Gamma (H_i \to \tau \tau)$,
as follows:
\beq
\Gamma (\phi_i\to \tau \mu) \, =\, (R^{\,\phi}_{\tau\mu})^2
                                   \, \Gamma (H_i \to \tau \tau)
\eeq
where
\beq
R^{\,\phi}_{\tau\mu}=\frac{g_{\phi \tau \mu}}{g_{\phi \tau \tau}}
\cong \frac{\sin(\alpha-\beta)}{\cos\alpha} \sqrt{\frac{m_{\mu}}{m_{\tau}}}
\, \tilde{\chi}_{23}
\eeq
Therefore, the Higgs boson branching ratio can be approximated as:
$Br(\phi_i \to \tau \mu)= (R^{\,\phi}_{\tau\mu})^2 \times 
Br(\phi_i \to \tau \tau)$. We calculated the branching fraction for
$h \to \tau\mu$,  and find that it reaches
values of order $10^{-2}$ in the THDM-III; for comparison, we notice
that in the MSSM case, even for large values of $\tan\beta$, 
one only gets  $Br(h \to \tau\mu)\simeq 10^{-4}$.

These values of the branching ratio enter into the domain of
detectability at hadron colliders (LHC), provided that the cross-section
for Higgs boson production were of order of the SM one.  Large values of
$\tan\beta$ are also associated with large b-quark Yukawa coupling,
which in turn can produce and enhancement on the Higgs boson production
cross-sections at hadron colliders, even for the heavier states $H^0$ and
$A^0$ either by gluon fusion or in the associated production of the Higgs
boson with b-quark pairs; some values are shown in table 1; these were
obtained using HIGLU \cite{hspira}. Thus, even the heavy Higgs bosons of
the model could be detected through this LFV mode.

\bigskip

\begin{center}
\begin{tabular}{|c|c|c|c|}
\hline
$m_{H,A}$ [GeV] & $\sigma^H_{gg}$ [pb] & $\sigma^A_{gg}$ [pb] &
 $\sigma^H_{bb}$ [pb] ($\simeq \sigma^A_{bb}$) \\
\hline
 150   & 126.4 (492.6) & 129.1 (525.) & 200 (800) \\
\hline
 200   & 29.5  (114.3) & 29.1 (120.)  & 100 (400) \\
\hline
 300   & 3.6   (13.5)  & 3.15 (13.6)  & 20 (80)  \\
\hline
 350   & 1.6   (5.9)   & 1.2  (5.6)   & 12 (48)  \\
\hline
 400   & 0.75  (2.75)  & 0.73 (2.8)   & 8 (32)  \\
\hline
\end{tabular}
\end{center}
\noindent
{Table\,1. Cross-section for Higgs boson production at LHC, through gluon
fusion ($\sigma^{H,A}_{gg}$) and in association with $b\bar{b}$ quarks,
($\sigma^{H,A}_{bb}$), for $\tan\beta=30$ (60).}

\bigskip

For instance, for $m_{H,A}=150$ GeV and $\tan\beta=30 (60)$
the cross-section through gluon fusion at LHC is about 
126.4 (492.6) pb \cite{hspira}, then with 
$Br(H\to \tau \mu) \simeq 10^{-2} (10^{-3})$ and
an integrated luminosity of $10^{5} \, pb^{-1}$,
LHC can produce about $10^5 (10^4)$ LFV Higgs boson events.
In Ref. \cite{htaumubkd} it was proposed a series of cuts to reconstruct
the hadronic and electronic tau decays from $h\to \tau\mu$ and separate the
signal from the backgrounds, which are dominated by Drell-Yan tau pair
and WW pair production.
According to these studies \cite{htaumubkd}, even SM-like cross 
sections and $m_\phi \simeq 150 \, GeV$, one coud detect at LHC
the LFV Higgs boson decays with a branching ratio of order $8 \times 10^{-4}$,
which means that our signal is clearly detectable.

\section{Conclusions}

We have studied in this paper the $f f' \phi^0$ couplings that
arise in the THDM-III, using a Hermitic four-texture form for the fermionic
Yukawa matrix. Because of this, although the $f f' \phi^0$
couplings are complex, the CP-properties of $h^0, H^0$ (even) and $A^0$
(odd) remmain valid.

We have derived bounds on the parameters of the model, using current
experimental bounds on LFV and FCNC transitions. One can say that the
present bounds on the couplings $\chi_{ij}$'s still allow the possibility
to study interesting direct flavor violating Higgs boson signals at future
colliders, provided one takes not too large values of $\tan\beta$
($\lsim 15$).

In particular, the LFV  couplings of the neutral Higgs bosons, can lead
to new discovery signatures of the Higgs boson itself. For instance,
the branching fraction for $h^0 \to \tau\bar{\mu}(\bar{\tau}\mu)$ can be
as large as $10^{-5}$, while $Br(h \to b\bar{s}(\bar{b}s))$ is also about
$10^{-4}$. These LFV Higgs modes complement the modes $B^0\to\mu\mu$,
$\tau \to 3\mu$, $\tau\to\mu\gamma$ and $\mu\to e\gamma$, as probes of
flavor violation in the THDM-III, which could provide key insights into
the form of the Yukawa mass matrix.

Thus, the coming generation of colliders will provide a decisive test of
the Yukawa sector of the SM and its extensions, as well as other properties
of the gauge-Higgs sector \cite{lista4}

\bigskip

\noindent{\bf Acknowledgments.}

\noindent J.L.D.-C. and A.R. would like to thank Sistema Nacional de
Investigadores (Mexico) for financial support, and the Huejotzingo
Seminar for inspiration. R.N.-P. acknowledges financial support from CONACYT
(Mexico). This research was supported in part by CONACYT (Mexico).

\newpage

\begin{center}
{\bf Figure Captions}
\end{center}

\noindent{\bf Fig. 1}: The upper bound $(\chi_{sb})_{u.\,b.}^
{b \to s \, \gamma}$ as a function of $\tan\beta$, for $\alpha = \beta$,
$\alpha = \beta - \pi /4$, $\alpha = \beta - \pi /3$,
with $\Delta Br(b \to s \, \gamma) < 10^{-5}$, taking  $m_{h^0} = 120$
$GeV$ and $\chi_{bb} = 0$.

\bigskip

\noindent{\bf Fig. 2}: The upper bound $(\chi_{sb})_{u.\,b.}^
{b \to s \, \gamma}$ as a function of $\tan\beta$, for $m_{h^0} = 80$
$GeV$, 120 $GeV$, 160 $GeV$, with $\Delta Br(b \to s \, \gamma) < 10^{-5}$,
taking  $\alpha = \beta - \pi /4$ and $\chi_{bb} = 0$.

\bigskip

\noindent{\bf Fig. 3}: The upper bound $(\chi_{sb})_{u.\,b.}^
{b \to s \, \gamma}$ as a function of $\tilde{\chi}_{bb}$, for
$\tan\beta = 5$, 25, 50, with $\Delta Br(b \to s \, \gamma) < 10^{-5}$,
taking $m_{h^0} = 120$ $GeV$ and $\alpha = \beta - \pi /4$.

\bigskip

\noindent{\bf Fig. 4}: The upper bound $(\chi_{sb})_{u.\,b.}^
{B^0_s \to \mu \mu}$ as a function of $\tan\beta$, for $\alpha = \beta$,
$\alpha = \beta - \pi /4$, $\alpha = \beta - \pi /3$,
with $\Gamma(B^0_{s} \to \mu^- \mu^+) < 8.7 \times 10^{-19}$ $GeV$,
taking  $m_{h^0} = 120$ $GeV$ and $\chi_{\mu\mu} = 0$.

\bigskip

\noindent{\bf Fig. 5}: The upper bound $(\chi_{sb})_{u.\,b.}^
{B^0_s \to \mu \mu}$ as a function of $\tan\beta$, for $m_{h^0} = 80$ $GeV$,
120 $GeV$, 160 $GeV$, with $\Gamma(B^0_{s} \to \mu^- \mu^+) < 8.7 \times
10^{-19}$ $GeV$, taking  $\alpha = \beta - \pi /4$ and $\chi_{\mu\mu} = 0$.

\bigskip

\noindent{\bf Fig. 6}: The upper bound $(\chi_{sb})_{u.\,b.}^
{B^0_s \to \mu \mu}$ as a function of $\tilde{\chi}_{\mu\mu}$, for
$\tan\beta = 5$, 25, 50, with $\Gamma(B^0_{s} \to \mu^- \mu^+) < 8.7 \times
10^{-19}$ $GeV$, taking $m_{h^0} = 120$ $GeV$ and $\alpha = \beta - \pi /4$.

\bigskip

\noindent{\bf Fig. 7}: Branching ratios for all the relevant decay modes
that are allowed
kinematically for $80 \, GeV < m_{h^0} < 160 \, GeV$; taking $\alpha =
\beta - 3 \pi/8$ with $\tan\beta = 2$; assuming $\tilde{\chi}_{ij} = 0.1$ for
$i = j$ and $i \neq j$.

\bigskip

\noindent{\bf Fig. 8}: Same as in Fig. 7, but for $\tan\beta = 2.61$

\bigskip

\noindent{\bf Fig. 9}: Same as in Fig. 7, but for $\tan\beta = 5$

\bigskip

\noindent{\bf Fig. 10}: Same as in Fig. 7, but for $\tan\beta = 15$

\bigskip

\noindent{\bf Fig. 11}: Same as in Fig. 7, but for $\tan\beta = 50$

\bigskip

\noindent{\bf Fig. 12}: Curves in the plane ($\alpha-\beta$) - $\tan\beta$
in which the coupling $b \bar{b} h^{0}$ vanishes, for $\tilde\chi_{bb}
= 0.01$, 0.1, 0.5, and 1.

\bigskip

\noindent{\bf Fig. 13}: Number of events $N^{\mu\mu\to\tau\mu}$
as a function of $\tan\beta$; taking $\chi_{\mu\tau} = 1 $,
for $s=m^2_{h^0}=(120 \, GeV)^2$, $\alpha = \beta$,
and yearly integrated luminosities $L_{year} =
0.1, \, 0.22, \, 1 \, fb^{-1}$ and beam energy resolutions of
$R = 0.003\%, \, 0.01\%, \, 0.1\%$, respectively.

\bigskip

\noindent{\bf Fig. 14}: Number of events $N^{\mu\mu\to\tau\mu}$
as a function of $\tan\beta$; taking $\chi_{\mu\tau} = 1 $,
for $s=m^2_{h^0}=(120 \, GeV)^2$, $\alpha = \beta - \pi /4$,
and yearly integrated luminosities $L_{year} =
0.1, \, 0.22, \, 1 \, fb^{-1}$ and beam energy resolutions of
$R = 0.003\%, \, 0.01\%, \, 0.1\%$, respectively.

\bigskip

\noindent{\bf Fig. 15}: Number of events $N^{\mu\mu\to\tau\mu}$
as a function of $\tan\beta$; taking $\chi_{\mu\tau} = 1 $,
for $s=m^2_{h^0}=(120 \, GeV)^2$, $\alpha = \beta - \pi /3$,
and yearly integrated luminosities $L_{year} =
0.1, \, 0.22, \, 1 \, fb^{-1}$ and beam energy resolutions of
$R = 0.003\%, \, 0.01\%, \, 0.1\%$, respectively.

\bigskip

\noindent{\bf Fig. 16}: Number of events $N^{\mu\mu\to\tau\mu}$
as a function of $\tan\beta$; taking $\chi_{\mu\tau} = 1 $,
for $s=m^2_{h^0}=(80 \, GeV)^2$, $\alpha = \beta - \pi /4$,
and yearly integrated luminosities $L_{year} =
0.1, \, 0.22, \, 1 \, fb^{-1}$ and beam energy resolutions of
$R = 0.003\%, \, 0.01\%, \, 0.1\%$, respectively.

\bigskip

\noindent{\bf Fig. 17}: Number of events $N^{\mu\mu\to\tau\mu}$
as a function of $\tan\beta$; taking $\chi_{\mu\tau} = 1 $,
for $s=m^2_{h^0}=(160 \, GeV)^2$, $\alpha = \beta - \pi /4$,
and yearly integrated luminosities $L_{year} =
0.1, \, 0.22, \, 1 \, fb^{-1}$ and beam energy resolutions of
$R = 0.003\%, \, 0.01\%, \, 0.1\%$, respectively.

\bigskip

\noindent{\bf Fig. 18}: Number of events $N^{\mu\mu\to\tau\mu}$
as a function of $\tan\beta$; taking
$\chi_{\mu\tau} =(\chi_{\mu\tau})_{u.\,b.}^ {\mu \to e \gamma}$
with $Br(\mu^+ \to e^+ \gamma) < 1.2 \times 10^{-11}$,
for $s=m^2_{h^0}=(120 \, GeV)^2$, $\alpha = \beta$,
and yearly integrated luminosities $L_{year} =
0.1, \, 0.22, \, 1 \, fb^{-1}$ and beam energy resolutions of
$R = 0.003\%, \, 0.01\%, \, 0.1\%$, respectively.

\bigskip

\noindent{\bf Fig. 19}: Number of events $N^{\mu\mu\to\tau\mu}$
as a function of $\tan\beta$; taking
$\chi_{\mu\tau} =(\chi_{\mu\tau})_{u.\,b.}^ {\mu \to e \gamma}$
with $Br(\mu^+ \to e^+ \gamma) < 1.2 \times 10^{-11}$,
for $s=m^2_{h^0}=(120 \, GeV)^2$, $\alpha = \beta - \pi /4$,
and yearly integrated luminosities $L_{year} =
0.1, \, 0.22, \, 1 \, fb^{-1}$ and beam energy resolutions of
$R = 0.003\%, \, 0.01\%, \, 0.1\%$, respectively.

\bigskip

\noindent{\bf Fig. 20}: Number of events $N^{\mu\mu\to\tau\mu}$
as a function of $\tan\beta$; taking
$\chi_{\mu\tau} =(\chi_{\mu\tau})_{u.\,b.}^ {\mu \to e \gamma}$
with $Br(\mu^+ \to e^+ \gamma) < 1.2 \times 10^{-11}$,
for $s=m^2_{h^0}=(120 \, GeV)^2$, $\alpha = \beta - \pi /3$,
and yearly integrated luminosities $L_{year} =
0.1, \, 0.22, \, 1 \, fb^{-1}$ and beam energy resolutions of
$R = 0.003\%, \, 0.01\%, \, 0.1\%$, respectively.

\bigskip

\noindent{\bf Fig. 21}: Number of events $N^{\mu\mu\to\tau\mu}$
as a function of $\tan\beta$; taking
$\chi_{\mu\tau} =(\chi_{\mu\tau})_{u.\,b.}^ {\mu \to e \gamma}$
with $Br(\mu^+ \to e^+ \gamma) < 1.2 \times 10^{-11}$,
for $s=m^2_{h^0}=(80 \, GeV)^2$, $\alpha = \beta - \pi /4$,
and yearly integrated luminosities $L_{year} =
0.1, \, 0.22, \, 1 \, fb^{-1}$ and beam energy resolutions of
$R = 0.003\%, \, 0.01\%, \, 0.1\%$, respectively.

\bigskip

\noindent{\bf Fig. 22}: Number of events $N^{\mu\mu\to\tau\mu}$
as a function of $\tan\beta$; taking
$\chi_{\mu\tau} =(\chi_{\mu\tau})_{u.\,b.}^ {\mu \to e \gamma}$
with $Br(\mu^+ \to e^+ \gamma) < 1.2 \times 10^{-11}$,
for $s=m^2_{h^0}=(160 \, GeV)^2$, $\alpha = \beta - \pi /4$,
and yearly integrated luminosities $L_{year} =
0.1, \, 0.22, \, 1 \, fb^{-1}$ and beam energy resolutions of
$R = 0.003\%, \, 0.01\%, \, 0.1\%$, respectively.

\bigskip

\noindent{\bf Fig. 23}: Number of events $N^{\mu\mu\to b s}$
as a function of $\tan\beta$; taking $\chi_{sb} = 1 $,
for $s=m^2_{h^0}=(120 \, GeV)^2$, $\alpha = \beta$,
and yearly integrated luminosities $L_{year} =
0.1, \, 0.22, \, 1 \, fb^{-1}$ and beam energy resolutions of
$R = 0.003\%, \, 0.01\%, \, 0.1\%$, respectively.

\bigskip

\noindent{\bf Fig. 24}: Number of events $N^{\mu\mu\to b s}$
as a function of $\tan\beta$; taking $\chi_{sb} = 1 $,
for $s=m^2_{h^0}=(120 \, GeV)^2$, $\alpha = \beta - \pi /4$,
and yearly integrated luminosities $L_{year} =
0.1, \, 0.22, \, 1 \, fb^{-1}$ and beam energy resolutions of
$R = 0.003\%, \, 0.01\%, \, 0.1\%$, respectively.

\bigskip

\noindent{\bf Fig. 25}: Number of events $N^{\mu\mu\to b s}$
as a function of $\tan\beta$; taking $\chi_{sb} = 1 $,
for $s=m^2_{h^0}=(120 \, GeV)^2$, $\alpha = \beta - \pi /3$,
and yearly integrated luminosities $L_{year} =
0.1, \, 0.22, \, 1 \, fb^{-1}$ and beam energy resolutions of
$R = 0.003\%, \, 0.01\%, \, 0.1\%$, respectively.

\bigskip

\noindent{\bf Fig. 26}: Number of events $N^{\mu\mu\to b s}$
as a function of $\tan\beta$; taking $\chi_{sb} = 1 $,
for $s=m^2_{h^0}=(80 \, GeV)^2$, $\alpha = \beta - \pi /4$,
and yearly integrated luminosities $L_{year} =
0.1, \, 0.22, \, 1 \, fb^{-1}$ and beam energy resolutions of
$R = 0.003\%, \, 0.01\%, \, 0.1\%$, respectively.

\bigskip

\noindent{\bf Fig. 27}: Number of events $N^{\mu\mu\to b s}$
as a function of $\tan\beta$; taking $\chi_{sb} = 1 $,
for $s=m^2_{h^0}=(160 \, GeV)^2$, $\alpha = \beta - \pi /4$,
and yearly integrated luminosities $L_{year} =
0.1, \, 0.22, \, 1 \, fb^{-1}$ and beam energy resolutions of
$R = 0.003\%, \, 0.01\%, \, 0.1\%$, respectively.

\bigskip

\noindent{\bf Fig. 28}: Number of events $N^{\mu\mu\to b s}$
as a function of $\tan\beta$; taking
$\chi_{sb} =(\chi_{sb})_{u.\,b.}^ {b \to s \gamma}$
with $\Delta Br(b \to s \, \gamma) < 10^{-5}$,
for $s=m^2_{h^0}=(120 \, GeV)^2$, $\alpha = \beta$,
and yearly integrated luminosities $L_{year} =
0.1, \, 0.22, \, 1 \, fb^{-1}$ and beam energy resolutions of
$R = 0.003\%, \, 0.01\%, \, 0.1\%$, respectively.

\bigskip

\noindent{\bf Fig. 29}: Number of events $N^{\mu\mu\to b s}$
as a function of $\tan\beta$; taking
$\chi_{sb} =(\chi_{sb})_{u.\,b.}^ {b \to s \gamma}$
with $\Delta Br(b \to s \, \gamma) < 10^{-5}$,
for $s=m^2_{h^0}=(120 \, GeV)^2$, $\alpha = \beta - \pi /4$,
and yearly integrated luminosities $L_{year} =
0.1, \, 0.22, \, 1 \, fb^{-1}$ and beam energy resolutions of
$R = 0.003\%, \, 0.01\%, \, 0.1\%$, respectively.

\bigskip

\noindent{\bf Fig. 30}: Number of events $N^{\mu\mu\to b s}$
as a function of $\tan\beta$; taking
$\chi_{sb} =(\chi_{sb})_{u.\,b.}^ {b \to s \gamma}$
with $\Delta Br(b \to s \, \gamma) < 10^{-5}$,
for $s=m^2_{h^0}=(120 \, GeV)^2$, $\alpha = \beta - \pi /3$,
and yearly integrated luminosities $L_{year} =
0.1, \, 0.22, \, 1 \, fb^{-1}$ and beam energy resolutions of
$R = 0.003\%, \, 0.01\%, \, 0.1\%$, respectively.

\bigskip

\noindent{\bf Fig. 31}: Number of events $N^{\mu\mu\to b s}$
as a function of $\tan\beta$; taking
$\chi_{sb} =(\chi_{sb})_{u.\,b.}^ {b \to s \gamma}$
with $\Delta Br(b \to s \, \gamma) < 10^{-5}$,
for $s=m^2_{h^0}=(80 \, GeV)^2$, $\alpha = \beta - \pi /4$,
and yearly integrated luminosities $L_{year} =
0.1, \, 0.22, \, 1 \, fb^{-1}$ and beam energy resolutions of
$R = 0.003\%, \, 0.01\%, \, 0.1\%$, respectively.

\bigskip

\noindent{\bf Fig. 32}: Number of events $N^{\mu\mu\to b s}$
as a function of $\tan\beta$; taking
$\chi_{sb} =(\chi_{sb})_{u.\,b.}^ {b \to s \gamma}$
with $\Delta Br(b \to s \, \gamma) < 10^{-5}$,
for $s=m^2_{h^0}=(160 \, GeV)^2$, $\alpha = \beta - \pi /4$,
and yearly integrated luminosities $L_{year} =
0.1, \, 0.22, \, 1 \, fb^{-1}$ and beam energy resolutions of
$R = 0.003\%, \, 0.01\%, \, 0.1\%$, respectively.

\bigskip

\noindent{\bf Fig. 33}: $|K|$ as a
function of $\tan\beta$; taking $m_{h^0} = 150 \, GeV$,
$\alpha =\beta - 3 \pi/8$. Assuming $\tilde{\chi}_{ij} = 0.1$
for $i \neq j$ and: (a) $\tilde{\chi}_{ii} = 0.01$ (line A);
(b) $\tilde{\chi}_{ii} = 0.1$ (line B); (c) $\tilde{\chi}_{ii} = 0.5$
(line C); (d) $\tilde{\chi}_{ii} = 1$ (line D).

\bigskip

\noindent{\bf Fig. 34}: Same as in Fig. 33, but for $\alpha =\beta - \pi/4$

\bigskip

\noindent{\bf Fig. 35}: Same as in Fig. 33, but for $\alpha =\beta - \pi/8$

\end{document}